\documentclass[reprint,superscriptaddress,amsmath,amssymb,aps]{revtex4-1}
\usepackage{graphicx} 
\usepackage{datetime}
\usepackage{dcolumn} 
\usepackage{bm} 
\usepackage{xcolor}

\begin{document}
\title{Surface plasmon lifetime in metal nanoshells}
\author{Arman S. Kirakosyan}
\affiliation{Department of Physics, Jackson State University, Jackson, MS 39217 USA}
\author{Mark I. Stockman}
\affiliation{Department of Physics and Astronomy, Georgia State University, Atlanta, GA 30303 USA}
\author{Tigran V. Shahbazyan$^{1,}$}
\thanks{ \texttt{shahbazyan@jsums.edu}}

\begin{abstract} 
The lifetime of localized surface plasmon plays an important role in many aspects of plasmonics and its applications. In small metal nanostructures, the dominant mechanism of plasmon decay is size-dependent Landau damping. We performed quantum-mechanical calculations of Landau damping for the bright surface plasmon mode in a metal nanoshell with dielectric core. In contrast to the conventional model based on the electron surface scattering, we found that  the damping rate decreases as  the nanoshell thickness is reduced. The origin of this behavior is traced to the  spatial distribution of plasmon local field in the metal shell.  We also found that, due to the interference of electron scattering amplitudes from  the two nanoshell  metal surfaces, the damping rate exhibits pronounced  quantum beats  with changing shell thickness.
\end{abstract}
\maketitle

\section{Introduction}

Lifetime of localized surface plasmons (SP) in metal nanostructures is one of the fundamental problems in plasmonics that has been continuously addressed for about 50 years \cite{kreibig-book,halperin-rmp86,kresin-pr92,schatz-jpcb03,noguez-jpcc07}. The importance of this issue stems from one of the major objectives of plasmonics -- generation of extremely strong local fields at the nanoscale. The range of physical phenomena and applications related to this goal cuts across physics, chemistry, biology, and device applications. A small sample of examples includes plasmon-enhanced spectroscopies of molecules or semicinductor quantum dots near metal nanostructures, such as surface-enhanced Raman scattering (SERS) \cite{sers}, plasmon-enhanced fluorescence \cite{feldmann-prl02,novotny-prl06,sandoghdar-prl06,halas-nl07}, plasmon-assisted fluorescence resonance energy transfer (FRET)  \cite{lakowicz-jf03,andrew-science04,krenn-nl08,lunz-nl11} and plasmonic laser (spaser) \cite{bergman-prl03,stockman-natphot08,noginov-nature09,zhang-nature09,gwo-science12,odom-natnano15}. High Ohmic losses in bulk metal due to strong electron-phonon interactions impose limitations on the quantum yield of metal-based plasmonic devices, which can, to some extent, be remedied by reducing the metal component size.

However, at the length scale below $\sim 10$ nm,  new limitations on the SP  lifetime and, consequently, on quantum yield arise due to the quantum-size effects \cite{kreibig-book}. Among those, the most important is the Landau damping (LD) of SP -- decay of SP into the Fermi sea electron-hole pair \cite{klar-prl98,mulvaney-prl02,halas-prb02,klar-nl04,vallee-prl04,halas-nl04,hartland-pccp06,vallee-nl09,vanduyne-jpcc12,vallee-nl13,schatz-nl15}. This process has been recently suggested as an efficient way  of hot carriers excitation in plasmon-based photovoltaic devices \cite{park-nl11,melosh-nl11,halas-nc13,link-acsnano13,lian-nl13,fabrizio-nn13,clavero-np14,brongersma-nl14,halas-nc15,wang-nc15,nordlander-nn15}. Starting with the pioneering work of Kawabata and Kubo \cite{kawabata-jpsj66} for a spherical nanoparticle (NP),  quantum-mechanical  calculations of LD rate were performed, using random phase approximation (RPA)  \cite{kawabata-jpsj66,lushnikov-zp74,schatz-jcp83,barma-pcm89,yannouleas-ap92,eto-srl96,uskov-plasmonics13,khurgin-oe15} or density functional theory (DFT) \cite{jalabert-prb02,jalabert-prb05,yuan-ss08,vallee-jpcl10,lerme-jpcc11,dionne-nature12,li-njp13,nordlander-acsnano14} methods, for several  NP  shapes. Excitation of an electron-hole pair with large optical frequency requires momentum relaxation to satisfy the energy and momentum conservation laws which, in small systems, can take place via the electron surface scattering. Based on this picture, it was suggested \cite{kreibig-zp75,ruppin-pss76,schatz-cpl83,schatz-jcp03,moroz-jpcc08} that  the SP LD rate in \textit{any} small system should have the form
\begin{align}
\label{ld_old}
&
\gamma_{s} = A\, \dfrac{v_{F}}{L},
\end{align}
where $v_{F}$ is the electron Fermi velocity (hereafter we set $\hbar=1$)  and $L$ is the    effective  mean free path of ballistic electrons confined  in a hard-wall potential well, while the  phenomenological constant $A$, measured in the range $0.3-1.5$ \cite{kreibig-book}, accounts for surface potential, electron spillover, and dielectric environment effects. Note that, for nonspherical NPs, the SP damping by interband excitations can complicate the LD size dependence. For example, absorption spectra for gold nanorods \cite{mulvaney-prl02,hartland-pccp06} and nanoshells \cite{halas-prb02,klar-nl04} show overall narrowing of the SP resonance that is redshifted  away from the interband transitions onset. At the same time, recent systematic studies of scattering spectra of single silver nanoprisms \cite{vanduyne-jpcc12}, gold nanorods \cite{vallee-nl13}, and gold nanodisks \cite{schatz-nl15} revealed significant discrepancies with Eq. (\ref{ld_old}), while no size-dependence was detected for the SP resonance width of single gold nanoshells \cite{halas-nl04}, implying that LD is shunted by the bulk SP damping  even for relatively thin  shells.

There is also a physical argument that renders Eq.~(\ref{ld_old}) invalid for nanostructures of general shape. Indeed, the rate of electron-pair  excitation by the SP local field  must be  sensitive to the field distribution in the NP. Note that for a solid sphere, the dipole SP electric field in  the NP is uniform and size independent, which is the reason  Eq.~(\ref{ld_old}) holds well for spherical NPs in a very wide size range \cite{kreibig-book}. However, in the general case,  the local field distribution  depends strongly on NP size or shape, so that the simple picture implied by  Eq.~(\ref{ld_old})  fails. Below we demonstrate  that the effect of field distribution  leads to a drastically different size and shape dependence of the LD decay rate in a nanostructure than that implied by Eq.~(\ref{ld_old}).

In this paper, we present a quantum-mechanical calculation of the LD rate for bright SP modes in  a metal nanoshell  (NS) with a dielectric core. We find that, with decreasing NS thickness $d$, the LD rate decreases as well,  in sharp contrast to the surface scattering model \cite{kreibig-zp75,ruppin-pss76,schatz-cpl83,schatz-jcp03,moroz-jpcc08}  predicting an increase of $\Gamma$ as the effective mean free path is reduced. Furthermore, for small overall NS sizes, the SP LD rate exhibits \textit{quantum beats} as a function of shell thickness caused by the interference between electron scattering amplitudes from the inner and outer NS boundaries. 

The paper is organized as follows. In Sec. \ref{sec:general} we outline our approach and present a formal expression for the LD rate in terms of the SP eigenmodes.  In Sec. \ref{sec:pleig} we describe the plasmon eigenmodes in metal NS with dielectric core and evaluate the NS internal energy. In Sec. \ref{sec:q}, we evaluate the power dissipated through electron-hole excitation by the SP eigenmodes. The calculated LD rates are discussed  in Sec. \ref{sec:disc}, and Sec. \ref{sec:conc} concludes the paper.

%
\section{Surface plasmon Landau damping rate in composite metal-dielectric nanostructures}
\label{sec:general}

In this section we outline our approach for calculations of the plasmon  damping rate in a composite metal-dielectric structure embedded in a dielectric medium. We assume that the structure is characterized by  dielectric function of the form $\varepsilon (\omega,\bm{r})=\varepsilon' (\omega,\bm{r})+i\varepsilon'' (\omega,\bm{r})$  and the retardation effects are unimportant. In the quasistatic case, the plasmon eigenmodes, labeled by $n$ here, are determined by the Gauss's law
\begin{align}
\label{maxwell}
\nabla \cdot \left [\varepsilon' (\omega_{n},{\bf r}) {\bf E}_{n}\right ]=0, 
\end{align}
where $\omega_{n}$ is the eigenfrequency, ${\bf E}_{n}=-\bm{\nabla} \Phi_{n}$ is the mode local field, and $\Phi_{n}$ is the potential. In the following we assume that only the  metal dielectric function $\varepsilon_{m}(\omega)=\varepsilon'_{m}(\omega)+i\varepsilon''_{m}(\omega)$ is complex and dispersive. The decay rate of a plasmon mode is given by \cite{shahbazyan-16}
\begin{equation}
\label{LD}
\Gamma_{n} =  Q_{n}/U_{n},
\end{equation}
where $U_{n}$ is the mode energy  \cite{landau},
\begin{align}
\label{energy-LL}
U_{n}=\! \int \! \frac{dV}{16\pi} \frac{\partial (\omega_{n}\varepsilon')}{\partial \omega_{n}}  |\textbf{E}_{n}|^{2}=\frac{\omega_{n}}{16\pi} \frac{\partial \varepsilon'_{m}}{\partial \omega_{n}}\! \int\! dV_{m} |\textbf{E}_{n}|^{2},
\end{align}
and $Q_{n}$ is the mode dissipated power
\begin{equation}
\label{power}
Q_{n}=\frac{\omega_{n}}{2}\,\text{Im}\int dV \textbf{E}_{n}^{*}\cdot \textbf{P}_{n},
\end{equation}
where $\textbf{P}_{n}$ is the polarization vector ($V_{m}$ stands for the metal volume). In the \textit{local} case, i.e.,  $\textbf{P}_{n}=\textbf{E}_{n}(\varepsilon-1)/4\pi$,   $Q$ is given by the usual expression \cite{landau}
\begin{equation}
Q_{n} =\frac{\omega_{n}\varepsilon''_{m}}{8\pi} \int dV_{m} |\textbf{E}_{n}|^{2},
\end{equation}
which, together with the mode energy (\ref{energy-LL}), yields the  standard plasmon  damping rate \cite{stockman},
\begin{equation}
\label{rate-main}
\Gamma_{n}=2\varepsilon''_{m}\left (\frac{\partial \varepsilon'_{m}}{\partial \omega_{n}}\right )^{-1}.
\end{equation}
For the Drude form of metal dielectric function, $\varepsilon_{m}=\varepsilon_{i}-\omega_{p}^{2}/\omega(\omega+i\gamma)$, where $\varepsilon_{i}$ is a weakly-dispersive interband contribution, $\omega_{p}$ is the bulk plasmon frequency and $\gamma$ is the scattering rate, one obtains $\gamma_{n}=\gamma$ for all modes.

The surface contribution $Q_{n}^{s}$ originates from the generation of electron-hole pairs by the plasmon local field near metal-dielectric interfaces, and can be included in Eq. (\ref{power})   by relating the polarization vector $\textbf{P}_{n}(\bm{r})$ to the microscopic electron polarization operator $P(\omega;\bm{r},\bm{r}')$ via  the induced charge density: $\rho(\bm{r})=\int d{\bm r}'P(\bm{r}, \bm{r}') \Phi(\bm{r}')=-\bm{\nabla}\cdot {\bf P}(\bm{r})$ \cite{shahbazyan-16}. Integrating Eq. (\ref{power}) by parts, we obtain
\begin{equation}
\label{power1}
Q_{n}^{s}=\frac{\omega_{n}}{2}\,\text{Im}\int dV dV' \Phi_{n}^{*}(\bm{r})P(\omega_{n};\bm{r},\bm{r}') \Phi_{n}(\bm{r}').
\end{equation}
In the first-order, $Q_{n}^{s}$ is obtained within RPA  as \cite{mahan-book}
\begin{equation}
\label{power-rpa}
Q_{n}^{s}=\pi \omega_{n}\sum_{\alpha\alpha'}|\langle \alpha'|\Phi_{n}|\alpha\rangle|^{2}\delta(\epsilon_{\alpha}-\epsilon_{\alpha'}+\omega_{n}),
\end{equation}
where $\langle \alpha'|\Phi_{n}|\alpha\rangle=\int dV_{m}\psi_{\alpha'}^{*}\Phi_{n}\psi_{\alpha}$ is the transition matrix element between electron state $\psi_{\alpha}$ with energy $\epsilon_{\alpha}$ \textit{below} the Fermi level $E_{F}$ and electron state $\psi_{\alpha'}$ with energy $\epsilon_{\alpha'}$ \textit{above} the Fermi level under the perturbation $\Phi_{n}$ (factor 2 due to the spin degeneracy is included). Note that often in the literature, the plasmon surface-assisted decay rate $\Gamma_{n}^{s}$ is  identified with the first-order transition probability rate, similar to Eq. (\ref{power-rpa}) (up to the factor $\omega_{n}/2$); it must be emphasized that, in a system with dispersive dielectric function, the accurate expression is $\Gamma_{n}^{s}=Q_{n}^{s}/U_{n}$ \cite{shahbazyan-16}. In the rest of this paper, this expression will be used to calculate the SP damping rate  in a metal NS.

\section{Plasmon modes in  metal nanoshells with dielectric core}
\label{sec:pleig}

\begin{figure}
\centering
\includegraphics[width=\columnwidth]{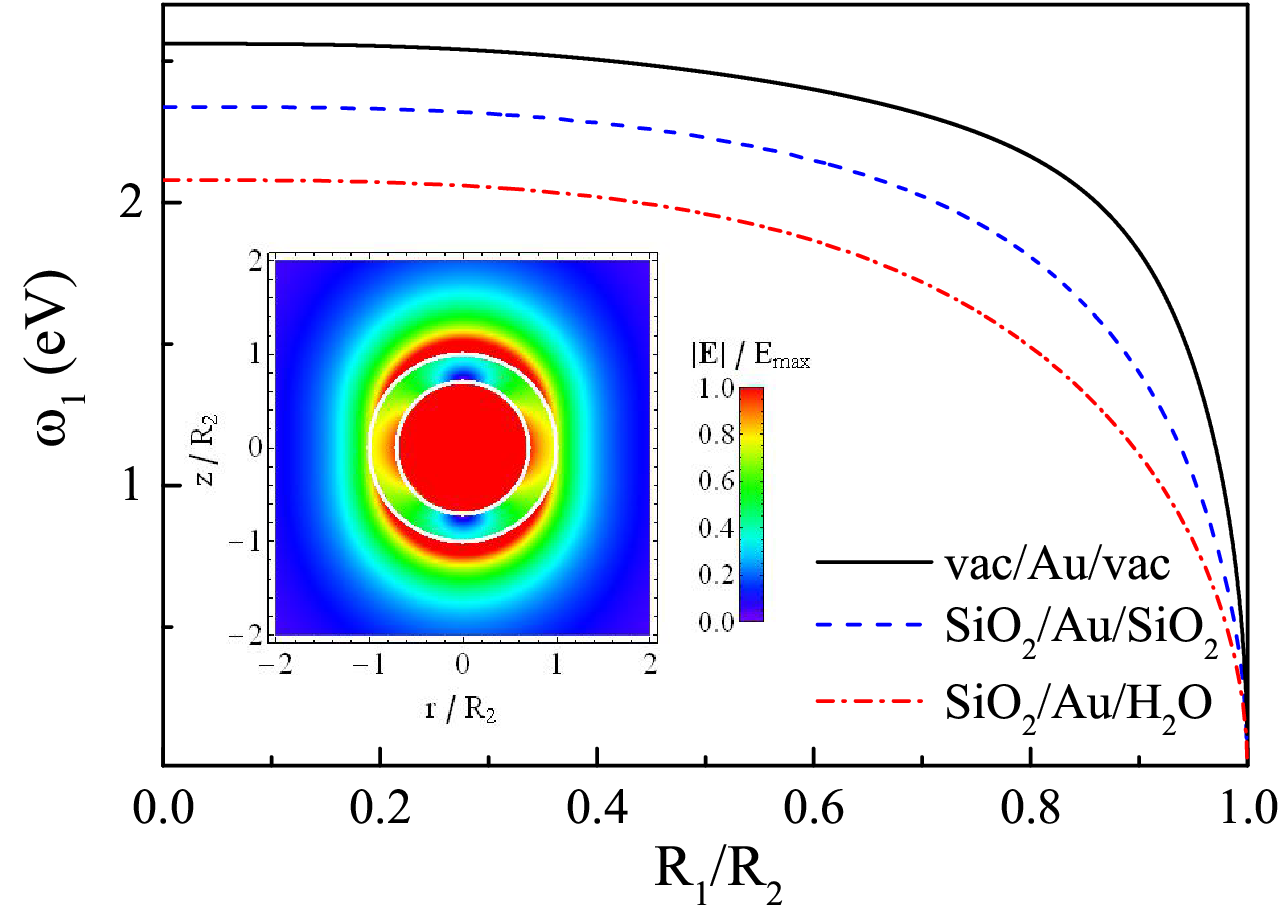}
\caption{
Frequency of bright dipole plasmon mode in gold NS with various core and outside dielectrics is  plotted vs. NS aspect ratio. Inset: Electric field distribution for $SiO_{2}/Au/H_{2}O$ NS with aspect  ratio $R_{1}/R_{2} = 0.7 $.}
\label{fig:spspectrum}
\end{figure}

Here we collect the relevant formulas for plasmonic eigenstates in a spherical NS with inner and outer radii $ R_{1} $ and $ R_{2} $, respectively, and core dielectric constant $\varepsilon_{c}$,   in a medium with dielectric constant $\varepsilon_{d}$ (see inset in Fig. \ref{fig:spspectrum}). In the quasistatic limit, the plasmonic eigenfunctions in each region have the form $\Phi_{LM}(\bm{r})=\Phi_{L}^{(i)}(r)Y_{LM}(\hat{\bm{r}})$, where  $r$ and $\hat{\bm{r}}$ are the magnitude and orientation of the radius vector with the origin at  NS center, $i=(c,m,d)$ denotes core, metal and outside dielectric regions, respectively, and $Y_{LM}(\hat{\bm{r}})$ are spherical harmonics. In each region, the  eigenfunctions are superpositions of two independent solutions of Laplace equation in spherical coordinates, $r^{L}Y_{LM}(\hat{\bm{r}})$ and $r^{-L-1}Y_{LM}(\hat{\bm{r}})$. The equation for eigenvalues is obtained by imposing standard boundary conditions on the radial part of potentials, $\Phi_{L}^{(i)}(r)$, and radial component of electric field, $E_{L}^{(i)}(r)=-\partial \Phi_{L}^{(i)}(r)/\partial r $, as
\begin{equation}
\label{eigenvalues}
\tilde{\varepsilon}_{cm}\tilde{\varepsilon}_{md}+L(L+1) \varepsilon_{cm}\varepsilon_{md}\kappa^{2L+1} = 0,
\end{equation}
where $ \kappa = R_{1}/R_{2} $ is the NS aspect ratio, and we denoted  $\varepsilon_{\alpha \beta} =  \varepsilon'_{\alpha} -\varepsilon'_{\beta}$ and $\tilde{\varepsilon}_{\alpha \beta} =  L \varepsilon'_{\alpha} + (L+1) \varepsilon'_{\beta} $. The plasmon frequencies are obtained by solving Eq. (\ref{eigenvalues}) for the real part of metal dielectric function,
\begin{align}
\label{eps_ns}
&
\varepsilon_{m}^{\prime}(\omega_{L}) = - \dfrac{\mu_{L}}{2}  \pm \sqrt{\dfrac{\mu^{2}_{L}}{4} -\varepsilon_{c} \varepsilon_{d}}, 
\nonumber\\&
\mu_{L} = \frac{(2L+1)}{L (L+1)}\frac{\tilde{\varepsilon}_{cd}}{(1-\kappa^{2L+1})}  - \varepsilon_{c} - \varepsilon_{d},
\end{align}
where alternating $(\pm)$ sign correspond to bright and dark plasmon modes, respectively. The bright plasmon spectrum matches that of  a solid NP plasmon  in the  $\kappa=0$ limit: $\tilde{\varepsilon}_{md} =  L \varepsilon'_{m} + (L+1) \varepsilon_{d} =0$.   The higher frequency dark plasmon mode couples weakly to the external fields and will not be considered here. 

In Fig. \ref{fig:spspectrum},  we show the dependence of bright plasmon mode frequency $\omega_{1}$ (for $L=1$) on aspect ratio $\kappa=R_{1}/R_{2}$ of Au NS with several choices of core and outside dielectrics. In all numerical calculations, the experimental dielectric function for gold as well as for core and outside dielectrics were used \cite{etchegoin-jcp06}. With decreasing shell thickness,  after a prolonged plateau for $\kappa$ up to approximately 0.5-0.7 (depending on dielectric content), the frequency develops a redshift. The inset shows electric field distribution for the dipole plasmon mode oscillating along the $z$ axis in a NS with $\kappa=0.7$. Note that, in a thin NS, the  electric field of bright plasmon is mainly concentrated \textit{outside} of the metal shell, in contrast to the field distribution in a  solid metal NP. 

The normalized  (dimensionless) radial eigenfunctions $\Phi_{L}(r)$ in core ($r<R_{1}$), shell ($R_{1}<r<R_{2}$), and outer dielectric ($r>R_{2}$) regions have the form
\begin{align}
\label{flc}
&
\Phi_{L}^{(c)}(r)
= (2L+1)\frac{\varepsilon_{m}^{\prime}\kappa^{L}}{\tilde{\varepsilon}_{cm}}\left (\frac{r}{R_{1}}\right )^{L},
\\ &
\label{flm}
\Phi_{L}^{(m)}(r) = \kappa^{L} \left (\frac{r}{R_{1}}\right )^{L} +
\frac{1}{L+1}\frac{\tilde{\epsilon}_{md}}{\epsilon_{md}}  \left (\frac{R_{2}}{r}\right )^{L+1},
\\ &
\label{fld}
\Phi_{L}^{(d)}(r)
=  \dfrac{2L+1}{L+1} \frac{\varepsilon_{m}^{\prime}}{\varepsilon_{md} }  \left (\frac{R_{2}}{r}\right )^{L+1} ,
\end{align}
and are continuous at the metal-dielectric interfaces,
\begin{align}
\label{interface-potential}
&
\Phi_{1L}\equiv \Phi_{L}^{(m)}(R_{1})
= (2L+1)\frac{\varepsilon_{m}^{\prime}\kappa^{L}}{\tilde{\varepsilon}_{cm}},
\\ &
\Phi_{2L} \equiv \Phi_{L}^{(m)}(R_{2})
=  \frac{2L+1}{L+1} \frac{\varepsilon_{m}^{\prime}}{\varepsilon_{md} }  .
\end{align}
The radial electric fields satisfy the standard boundary conditions, i.e., $\varepsilon_{\alpha}E_{L}^{(\alpha)}(r)$ is continuous, and take the following values at the interfaces (on the metal side)
\begin{align}
\label{interface-field}
&
E_{1L}\equiv E_{L}^{(m)}(R_{1})
=-\frac{L}{R_{1}}  \frac{\varepsilon_{c}}{{\varepsilon}'_{m}}\Phi_{1L},
\nonumber\\ &
E_{2L} \equiv E_{L}^{(m)}(R_{2})
=  \frac{L+1}{R_{2}} \frac{\varepsilon_{d} }{\varepsilon'_{m} }\Phi_{2L},
\end{align}
while their ratio at the interfaces is given by
\begin{equation}
\label{ratio}
q_{L} = E_{1L}/E_{2L}= -L \kappa^{L-1} \dfrac{\varepsilon_{md} \varepsilon_{c}}{\tilde{\varepsilon}_{cm} \varepsilon_{d}}.
\end{equation}
Note that the electric field orientations at the inner and outer interfaces (on the metal side) are opposite.

Using the above   eigenfunctions, the plasmon mode  energy can be straightforwardly calculated  from  Eq. (\ref{energy-LL}). Since the eigenfunctions are harmonic functions inside each region, the integral in Eq. (\ref{energy-LL}) reduces to the boundary terms, and, using the relations (\ref{interface-field}) between fields and potentials at the interface, we obtain
\begin{equation}
\label{energy-electric}
U_{L}=\frac{|\varepsilon'_{m}|\omega_{L}}{16\pi}
\frac{\partial \varepsilon'_{m}}{\partial\omega_{L}}
\left [\frac{R_{1}^{3}}{L\epsilon_{c}}E_{1L}^{2}+\frac{R_{2}^{3}}{(L+1)\epsilon_{d}}E_{2L}^{2}\right ].
\end{equation}
%

\begin{figure}
\centering
\includegraphics[width=\columnwidth]{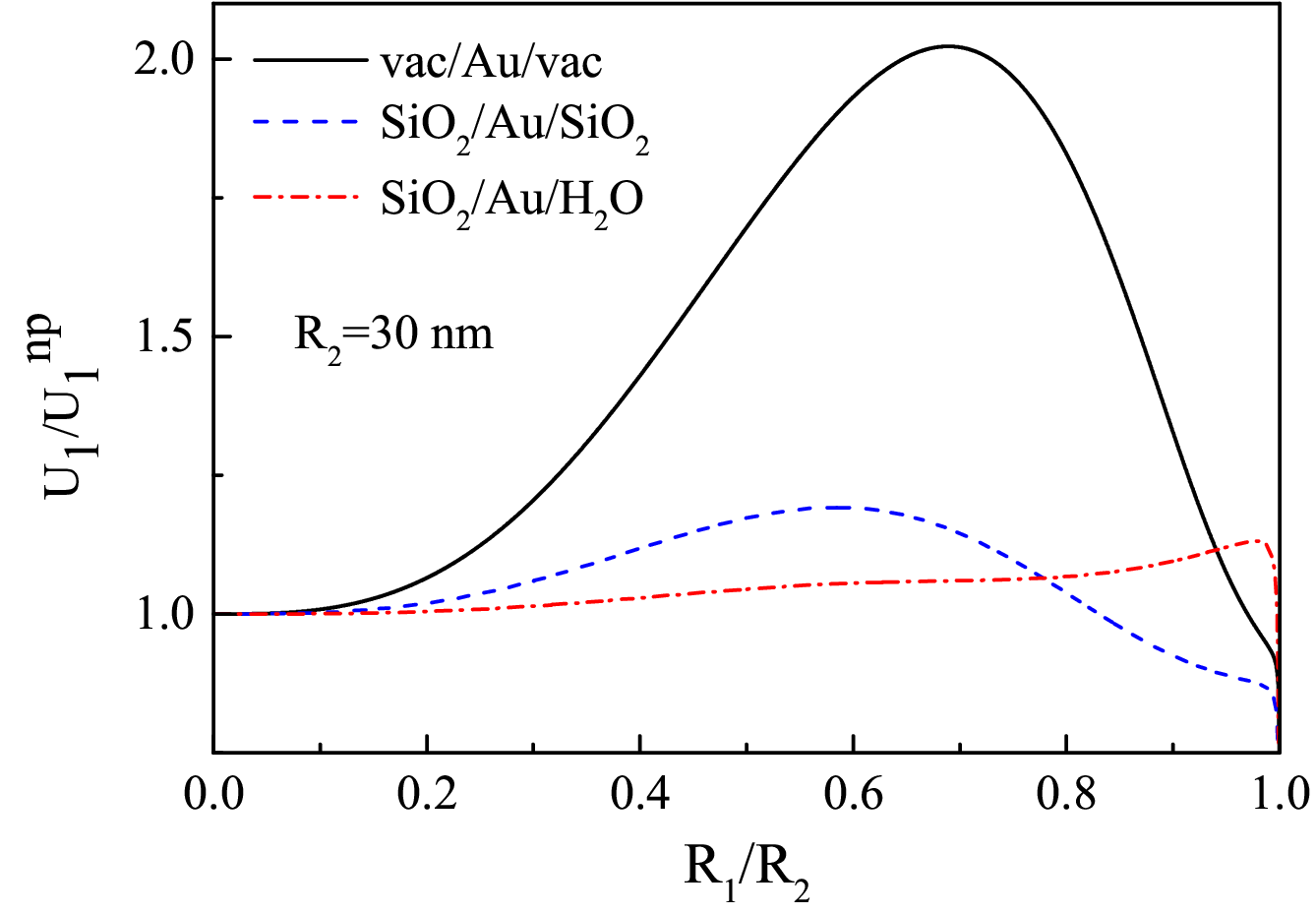}
\caption{
Normalized energy of bright dipole plasmon modes in gold NS with various core and outside dielectrics is plotted vs. NS aspect ratio. 
}
\label{fig:energy}
\end{figure}

The aspect ratio dependence of the  bright dipole  plasmon   energy $U_{1}$ normalized to solid NP plasmon energy $U_{1}^{np}$ with the same overall size is plotted in Fig.~\ref{fig:energy}. The NS mode energy depends strongly on core and outside dielectrics, but is largely  comparable to that for a solid NP. This is due to a somewhat similar distribution of the surface charges for bright NS plasmon and solid NP plasmon modes: In both cases,  the opposite charges are located at different hemispheres  so the energy is proportional to the core-shell particle volume. In contrast, for dark modes (not shown here), the opposite charges are located at inner and outer boundaries, so the energy vanishes as the shell thickness decreases.
%
\section{Power dissipated by plasmon modes in nanoshells}
\label{sec:q}

We now turn to   calculation of dissipated power Eq. (\ref{power-rpa}) (we drop superscript $s$ in the following). We represent the NS confining potential as a three-dimensional quantum well with hard boundaries at $R_{1}$ and $R_{2}$ and amplitude $V_{0}$:  $V(r)=V_{0}\theta(r-R_{1})\theta(R_{2}-r)$. The role of realistic surface potential and nonlocal effects will be discussed later. The electron wave functions have the form $\psi_{nl}(r)Y_{lm}(\bm{r})$, where $n$, $l$ and $m$ and electron radial, angular momentum and magnetic numbers, respectively. Due to spherical symmetry, the angular part factorizes out and Eq. (\ref{power-rpa}) takes the form
\begin{eqnarray}
\label{Q}
Q_{L}=\pi\omega_{L}\sum\limits_{nn'll'}a_{ll'}^{L}|M_{nl,n'l'}^{L}|^2
\delta(\epsilon_{nl} - \epsilon_{n'l'}+\omega_{L}),
\end{eqnarray}
where $M_{nl,n'l'}^{L}=\langle nl | \Phi_{L}| n'l' \rangle$ is the radial transition matrix element and 
\begin{equation}
a_{ll'}^{L}=\frac{1}{2L+1}\sum\limits_{Mmm'}\left|\int d {\bm \Omega} Y_{L M}Y^{*}_{l m} Y_{l' m'} \right|^2
\end{equation}
is the angular contribution. The latter is nonzero only for $l=l'\pm L$, and for typical $l,l'\gg L$, can be approximated as  $a_{ll'}^{L}\approx \delta_{ll'}l/2\pi$.

The matrix element $\langle\alpha|\Phi_{LM}|\alpha'\rangle$ in  Eq. (\ref{power-rpa}) is dominated by the surface contribution, which can be obtained by first commuting twice the plasmon potential $\Phi_{LM}$ with the Hamiltonian,
\begin{align}
\langle\alpha|\Phi_{LM}|\alpha'\rangle
&=\frac{1}{\omega_{L}^{2}}\langle\alpha\left[H\left[H,\Phi_{LM}\right]\right]|\alpha'\rangle
\nonumber\\
&\approx\frac{1}{m\omega_{L}^{2}} \langle\alpha|\nabla \Phi_{LM}\cdot\nabla V|\alpha'\rangle,
\end{align}
which, after separating out the angular part, leads to the following expression for the radial matrix element,
\begin{align}
M_{nl,n'l'}^{L}=
\frac{V_{0}}{m\omega_{L}^{2}}\bigl [\psi_{nl}(R_{1})\psi_{n'l'}(R_{1}){\cal E}_{L}(R_{1})
\nonumber\\
-\psi_{nl}(R_{2})\psi_{n'l'}(R_{2}){\cal E}_{L}(R_{2})\bigr ].
\end{align}
Then, for infinitely high potential barrier ($V_{0}\rightarrow \infty$), matching the wave-functions across the well boundaries gives $\sqrt{2mV_{0}}\psi_{nl}(R_{i})\approx -\psi'_{nl}(R_{i})$ (here prime stands for the derivative), and the matrix element takes the form
\begin{align}
\label{matrix}
M_{nl,n'l'}^{L}=
\frac{1}{2m^{2}\omega_{L}^{2}}\bigl [\psi'_{nl}(R_{1})\psi'_{n'l'}(R_{1})E_{1L}
\nonumber\\
-\psi'_{nl}(R_{2})\psi'_{n'l'}(R_{2})E_{2L}\bigr ].
\end{align}
The first and second terms in the r.h.s. describe excitation, by the plasmon electric field, of a Fermi sea electron-hole pair accompanied by  momentum transfer to the inner and outer boundaries, respectively. Correspondingly, $Q_{L}$ can be decomposed as $Q_{L}=Q_{L}^{11}+Q_{L}^{22}-2Q_{L}^{12}$, where
\begin{align}
\label{Qij}
&Q_{L}^{ij}=
\frac{e^{2}}{8m^{4}\omega_{L}^{3}}E_{1L}E_{2L}
\\
&\times\!\sum\limits_{lnn'} \psi'_{nl}(R_{i})\psi'_{n'l}(R_{i})\psi'_{nl}(R_{j})\psi'_{n'l}(R_{j})\delta(\epsilon_{nl} - \epsilon_{n'l}+\omega_{L}),
\nonumber
\end{align}
and we used that $a_{ll'}^{L}\approx \delta_{ll'}l/2\pi$. 

Consider first the inner surface contribution, $Q_{L}^{11}$. For typical electron energies $\epsilon_{nl}\sim E_{F}$, we can adopt semiclassical approximation for the electron wave-functions:
\begin{equation}
\psi_{nl}(r)=\sqrt{\frac{4m}{p_{l}\tau_{l}}}\sin\int_{r}^{R_{2}}p_{l}dr,
~
p_{l} =\sqrt{2m\epsilon-\frac{(l+1/2)^{2}}{r^{2}}},
\end{equation}
where  $\tau_{l} (\epsilon)$ is the period of classical motion between two turning points. In this case, we find 
\begin{equation}
\label{psi1'}
\psi'_{nl}(R_{1})=-\sqrt{4mp_{l}(R_{1})/\tau_{l}}.
\end{equation}
Since the plasmon energy $ \omega_{L} $ is much larger than  the spacing $ \epsilon_{0}=v_{F}/d $  between the energy levels with adjacent $n$ (at fixed $ l $) in a spherical well,  the sums in Eq.~(\ref{Qij}) can be replaced by the integrals, $\sum_{n}\rightarrow \int d\epsilon \rho_{l}({\epsilon)}$  (with $\epsilon < E_{F}, \epsilon' > E_{F}$), where  $\rho_{l}(\epsilon)={\partial n}/{ \partial\epsilon_{nl}}$ is the  partial density of states related to the classical period as $\rho_{l}=\tau_{l}/2\pi$ (see Appendix). The result reads
\begin{equation}
\label{Q11}
Q_{L}^{11}=
\frac{E_{1L}^{2}}{2\pi^2m^{2}\omega_{L}^{3}}
\sum_{l} l \int_{E_{F}-\omega_{L}}^{E_{F}} 
d\epsilon p_{l}(\epsilon,R_{1})p_{l}(\epsilon+\omega_{L},R_{1}).
\end{equation}
Note that $\rho_{l}$ cancels out, i.e., the level spacing disappears from the result. In the energy integral, the integration variable is first shifted as  $\epsilon\rightarrow E_{F}+\epsilon-\omega_{L}/2$, where $\epsilon$ now changes in the interval $(-\omega_{L}/2, \omega_{L}/2)$, and then rescaled to $x=\epsilon/\omega_{L}$. The sum over $l$ is replaced by the integral restricted by maximal value $l\sim p_{F}R_{1}$ that is determined by the condition $p_{l}(\epsilon,R_{1})\geq 0$. After the change of variables to $s=l^{2}/(p_{F}R_{1})^{2}$, it contributes a factor proportional to the inner surface area. The result reads
\begin{eqnarray}
\label{Q11-final}
Q_{L}^{11}=
\frac{E_{F}^{2}R_{1}^{2}}{2\pi^2\omega_{L}^{2}}\, E_{1L}^{2}\,g\left(\omega/E_{F}\right),
\end{eqnarray}
where  $g(\xi)=2\int_{-1/2}^{1/2}dx\int ds f(\xi,x,s)$ with $f(\xi,x,s)=[\left (1+\xi x -s\right )^{2}-\xi^{2}/4]^{1/2}$, is a dimensionless function normalized to $g(0)=1$.

Turning to the outer surface term, $Q_{L}^{22}$, the main contribution into the r.h.s. of Eq.~(\ref{Qij}) comes from the terms with $p_{l}(\epsilon,R_{2})\geq 0$ [otherwise $\psi_{nl}(R_{2})$ are exponentially small]. In this case, we have 
\begin{equation}
\label{psi2'}
\psi'_{nl}(R_{2})=-(-1)^{n}\sqrt{4mp_{l}(R_{2})/\tau_{l}},
\end{equation}
where the sign factor $(-1)^{n}  $ accounts for the \emph{parity} of electron wave function with $n-1$ nodes between $R_{1}$ and $R_{2}$. The rest of the calculation is carried in a similar way, and the result,
\begin{eqnarray}
\label{Q22-final}
Q_{L}^{22}=
\frac{ E_{F}^{2}R_{2}^{2}}{2\pi^2\omega_{L}^{2}} \,E_{2L}^{2}\,g(\omega/E_{F}),
\end{eqnarray}
is proportional to the outer surface area.

Finally, consider now the interference term $Q_{L}^{12}$. Using Eqs. (\ref{psi1'}) and (\ref{psi2'}), we write
\begin{align}
\label{Q12}
Q_{L}^{12}=
\frac{2 E_{1L}E_{2L}}{m^{2}\omega_{L}^{3}}
&\sum_{lnn'}\frac{l(-1)^{n-n'}}{\tau_{l}(\epsilon_{nl})\tau_{l}(\epsilon_{n'l})}
\\
&\times F_{l}(\epsilon_{nl},\epsilon_{n'l})\delta(\epsilon_{nl} - \epsilon_{n'l}+\omega_{L}),
\nonumber
\end{align}
with $F_{l}(\epsilon,\epsilon')=\sqrt{p_{l}(\epsilon,R_{1})p_{l}(\epsilon',R_{1})p_{l}(\epsilon,R_{2})p_{l}(\epsilon',R_{2})}$.  As $\omega_{L}$ changes (e.g., with changing aspect ratio), the relative parity of  electron and hole states, separated by  energy $\omega_{L}$, changes too, leading to a different sequence of alternating signs in the sum in Eq.~(\ref{Q12}) which, in turn, results in oscillations of $Q_{L}^{12}$ (quantum beats). The number of states contributing into the sum in Eq.~(\ref{Q12}) is large, so that the oscillations  can be described   by substituting $(-1)^{n-n'}=\cos\pi(n-n')=\cos\Bigl[\pi\int_{\epsilon}^{\epsilon'}d\epsilon \rho_{l}(\epsilon)\Bigr]$. Then $Q_{L}^{12} $ takes the form
\begin{align}
\label{Q12-int}
Q_{L}^{12}=
\frac{E_{1L}E_{2L}}{2\pi^2m^{2}\omega_{L}^{3}}
&\sum_{l}l\int_{E_{F}-\omega_{L}}^{E_{F}} d\epsilon 
F_{l}\left (\epsilon,\epsilon+\omega_{L}\right )
\nonumber\\
&\times\cos\left [\pi \int_{\epsilon}^{\epsilon+\omega_{L}}d\epsilon' \rho_{l}(\epsilon')\right ] ,
\end{align}
where $l$ is restricted by the condition $p_{l}(\epsilon,R_{1})\geq 0$. Equation (\ref{Q12-int}) can be brought to the form
\begin{eqnarray}
\label{Q12-final}
Q_{L}^{12}=
\frac{e^{2} R_{1}^{2}E_{F}^{2}}{2\pi^2\omega_{L}^{2}}\, E_{1L}E_{2L}\,G(\omega_{L}/E_{F}),
\end{eqnarray}
where the dimensionless function $G(\xi)$ is rather cumbersome and is given in the Appendix. For thin nanoshells, $d/R_{2}\ll 1$, it can be evaluated analytically (see Appendix) and the result reads
\begin{equation}
\label{G-anal}
G(\xi)=-4\frac{\sin D}{D}\frac{\sin(\xi D/4)}{\xi D/4},
\end{equation}
where $D=\omega_{L}/\epsilon_{0}=\omega_{L}d/v_{F}$ is the ratio of plasmonic and electronic energy scales.

Putting all together, we finally obtain
\begin{eqnarray}
\label{Q-final}
Q_{L}=\frac{E_{F}^{2} R_{2}^{2}}{2\pi^2\omega_{L}^{2}}
\left (E_{2L}^{2}+\kappa^{2}E_{1L}^{2}-2\kappa^{2} E_{1L}E_{2L} G\right ).
\end{eqnarray}
The last term in $Q_{12}$ oscillates as a function of shell thickness $d$ due to the interference of electron scattering amplitudes from inner and outer NS boundaries.  These oscillations are, in fact, \textit{quantum beats} caused by the change, with $d$, of the number of electron levels with alternating parities within the plasmon energy  $\omega_{L}$ (i.e., the difference between numbers of even and odd states oscillates between 0 and 1). The oscillations period $2\pi v_{F}/\omega_{L}$ depends weakly on the shell thickness through dependence of $\omega_{L}$ on $\kappa$ (see Fig. \ref{fig:spspectrum}), and their amplitude slowly dies out with increasing $d$.

In fact, the quantum beats of $Q_{12}$ have a rather general origin. Indeed, excitation of an electron-hole pair with energy $\omega$ is accompanied by momentum transfer $p_{0}\sim \omega/v_{F}$ and occurs in a region with the size  $r_{0}\sim v_{F}/\omega$. Therefore, oscillations of the pair excitation rate with changing $D=d/r_{0}$ reflect the nonlocality of surface-scattering mechanism of momentum relaxation.

\begin{figure}[bt]
\centering
\includegraphics[width=\columnwidth]{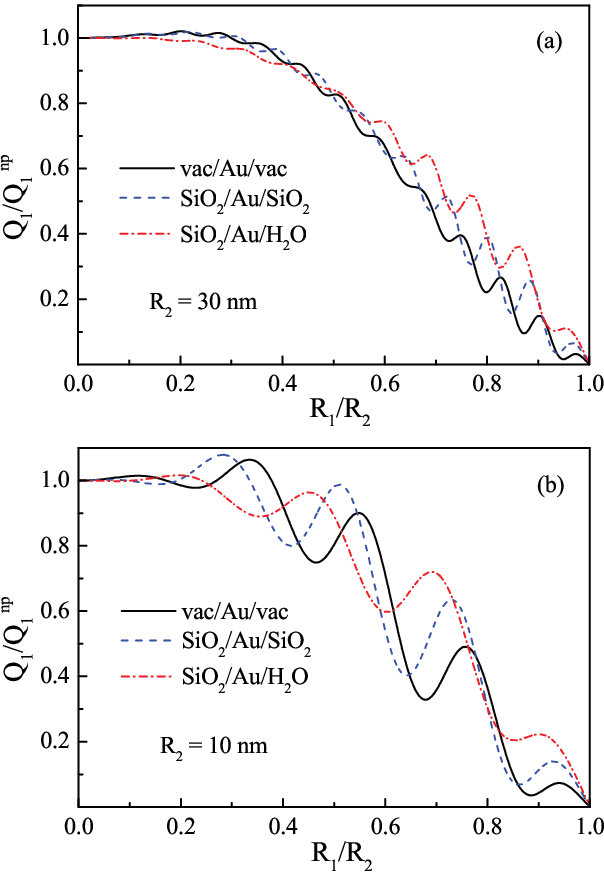}
\caption{
Normalized dissipated power by bright dipole plasmon modes in gold NS with various core and outside dielectrics is plotted vs. NS aspect ratio for (a) $R_{2}=30$ nm and (b) $R_{2}=10$ nm. 
}
\label{fig:power}
\end{figure}

In Fig. \ref{fig:power}, normalized dissipated power for the  bright dipole plasmon mode $Q_{1}$ is plotted vs.  aspect ratio $\kappa$ for  overall NS sizes $R_{2}=30$ nm and $R_{2}=10$ nm. Numerical calculations were performed using the full expression for  $G(\xi)$ given by Eq. (\ref{G}) in the Appendix. While  for larger NS with overall size $R_{2}=30$ nm, oscillations of $Q_{1}$ are relatively weak  [see Fig. \ref{fig:power}(a)], they become more pronounced for smaller NS ($R_{2}=10$ nm) [see Fig. \ref{fig:power}(b)]. Note that, for smaller $R_{2}$,  the same values of $\kappa$ correspond to smaller shell thicknesses. Another striking feature is the \textit{decrease} of dissipated power for $\kappa$ larger than 0.4. The reason for this behavior is that, with decreasing shell thickness, the local field is pushed outside the metal shell (see inset in Fig. \ref{fig:spspectrum}) which, in turn, leads to the reduction of the transition matrix element.

\section{Landau damping of plasmon modes in nanoshells}
\label{sec:disc}

The plasmon damping rate,  $\Gamma_{L}=Q_{L}/U_{L}$ with $Q_{L}$ and $U_{L}$ given by Eqs.~(\ref{Q-final}) and (\ref{energy-electric}), respectively, takes the form
\begin{equation}
\label{rate-general}
\Gamma_{L}=\frac{2\omega_{p}^{2}\gamma_{L}}{\omega_{L}^{3}} \left (\frac{\partial \varepsilon'_{m}}{\partial \omega_{n}}\right )^{-1},
\end{equation}
where
\begin{equation}
\label{ld-rate}
\gamma_{L}=
\frac{3v_{F}}{4R_{2}}\,
\frac{\varepsilon_{d} (L+1)}{|\varepsilon^{\prime}_{m}(\omega_{L})|}\,
\frac{1+\kappa^{2}q_{L}^{2}-2\kappa^{2}q_{L}G}{1+\kappa^{3}q_{L}^{2}(L+1)\varepsilon_{d}/L\varepsilon_{c}}
\end{equation}
is the LD rate. Here $q_{L}=E_{1L}/E_{2L}$ is the electric fields' ratio at the interfaces given by Eq. (\ref{ratio}). In deriving Eq.  (\ref{ld-rate}), we used the relation $\omega_{p}^{2}=4\pi n/m=4p_{F}^{3}/3\pi m$ (for $e=1$), where $n$ is the electron concentration.

Equations (\ref{rate-general}) and (\ref{ld-rate}) represent our central result. Apart from the dimensional factor $v_{F}/R_{2}$, the LD rate (\ref{ld-rate}) is determined by the ratio of plasmon local fields at the metal-dielectric interfaces $q_{L}$. The last factor describes the relative contribution of the NS interfaces and includes the interference correction. Importantly, comparison of Eqs.~(\ref{rate-general}) and (\ref{rate-main}) indicates that LD rate can be incorporated into the Drude scattering rate as $\gamma=\gamma_{0}+\gamma_{L}$, where $\gamma_{0}$ is the bulk scattering rate, so the full plasmon damping rate is still given by Eq.~(\ref{rate-main}), but with modified Drude dielectric function.

\begin{figure}[bt]
\centering
\includegraphics[width=\columnwidth]{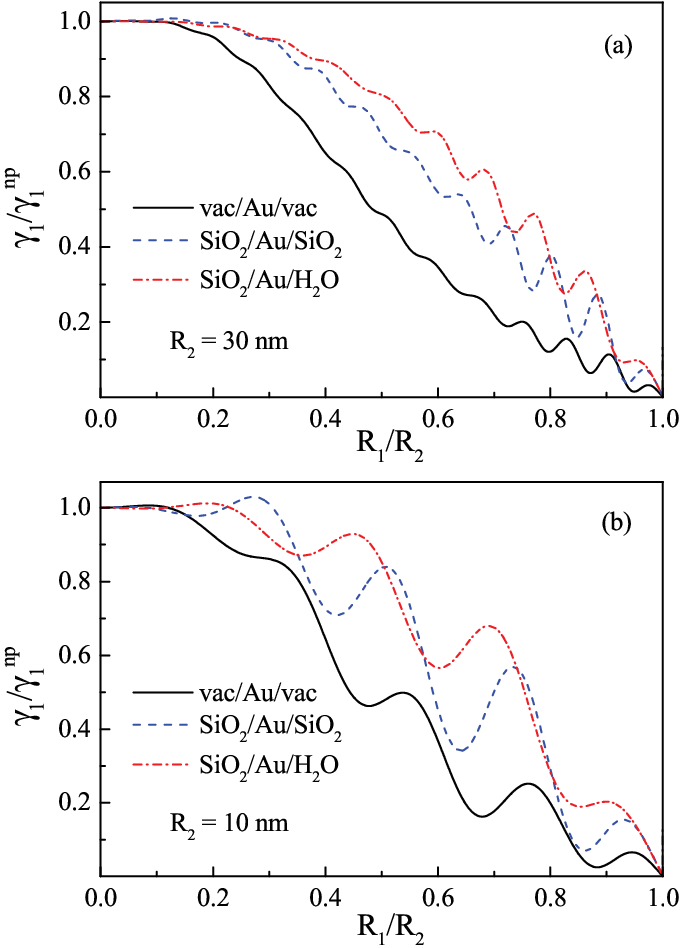}
\caption{
Normalized Landau damping rate for bright dipole plasmon modes in gold NS with various core and outside dielectrics is plotted vs. NS aspect ratio for (a) $R_{2}=30$ nm and (b) $R_{2}=10$ nm. 
}
\label{fig:ld}
\end{figure}
For solid NP ($\kappa=0$), the plasmon eigenfrequency is determined from $L\varepsilon_{m}(\omega_{L})+ (L+1)\varepsilon_{d}=0$, and we recover the LD rate of the $L$th mode in a spherical NP  \cite{kawabata-jpsj66,lushnikov-zp74,schatz-jcp83,barma-pcm89,yannouleas-ap92},
\begin{eqnarray}
\label{G-particle}
\gamma_{L}^{np}=\frac{3L}{4}\frac{v_{F}}{R_{2}}.
\end{eqnarray}
In thin NSs, the electric field is pushed out of the metal shell, leading to the reduction of electron-hole excitation rate. For thin NS ($d/R_{2}\ll 1$), the explicit dependence of the LD rate on the shell thickness is obtained  from Eq. (\ref{ld-rate}) as   (for $L=1$)
\begin{eqnarray}
\label{G-thin}
\gamma_{1}\approx \frac{3}{2}\frac{ v_{F}d}{ R_{2}^{2}} \left[1 - \dfrac{4\varepsilon_{c}\varepsilon_{d} }{\tilde{\varepsilon}^{2}_{cd}}(1-G)\right],
%
\end{eqnarray}
indicating a \textit{linear} dependence on the shell thickness.

In Fig.~\ref{fig:ld}, we show the calculated LD rate $\gamma_{1}$ for the bright dipole plasmon mode in gold NSs of overall sizes $R_{2}=30$ nm and $R_{2}=10$ nm  and several choices of core and outside dielectrics. The  rate shows approximately linear decrease with increasing $\kappa$ (i.e., decreasing $d$), consistent with Eq.~(\ref{G-thin}). The oscillations of $\gamma_{1}$ are quite pronounced for smaller overall NS size ($R_{2}=10$ nm) and could be observable for typical experimental range of aspect ratios (0.6-0.8) provided that NS overall size is sufficiently small, so that the LD  is not shunted by the bulk scattering. Note that these oscillations should be distinguished from those observed in solid NP \cite{jalabert-prb02,jalabert-prb05,li-njp13} due to size-quantization of the electron energy levels in a confined nanostructure, while here they are  quantum beats between electron scattering paths  from different NS interfaces.



\section{Conclusions}
\label{sec:conc}

In conclusion, let us discuss the role surface potential, dielectric environment and nonlocal effects near the metal surface on the plasmon LD that was extensively studied in solid  NPs \cite{jalabert-prb02,jalabert-prb05,yuan-ss08,vallee-jpcl10,lerme-jpcc11,dionne-nature12,li-njp13,nordlander-acsnano14}. These effects mainly affects the overall magnitude of LD rates, but plays no significant role in determining the LD dependence on the nanostructure shape which, according to our findings, is mainly determined by the local field ratio at the interfaces. Extensive theoretical  and experimental studies of spherical NPs indicate that surface effects mainly affect the phenomenological constant $A$ [see Eq.~(\ref{ld_old})], but the overall $1/R$ dependence of the LD rate is unchanged \cite{kreibig-book}. In fact, the important role of local fields in plasmon LD rate can explain the relatively wide range of measured  $A$ (0.3-1.5 \cite{kreibig-book}), which raised questions about the validity of the scattering model \cite{dionne-nature12}.  Indeed, as we mentioned in Sec. \ref{sec:q}, excitation of an \textit{e-h} pair by plasmon local field takes place in a surface layer of thickness $r_{0}\sim v_{F}/\omega$. For  $v_{F}\approx 1.4\times 10^{6}$ m/s  in Au and Ag, we have $v_{F}/\omega\approx 1$ nm for $\hbar\omega=1.0$ eV, i.e., for typical plasmon frequencies in the range 1.5-3.5  eV, the  layer thickness is just a few $\text{\AA}$. In a thin surface layer, the local fields are strongly affected by the electron spillover and surface roughness effects as well as by the dielectric environment, which can lead to large variations of overall LD rate magnitude for different samples and/or environments.  Within our approach, the constant $A$ can be estimated by computing the effect of the above factors on the local field,  which is, however, out of the scope of this paper.

In summary, we calculated the Landau damping rate of surface plasmons in metal nanoshells with dielectric core. We found that the damping rate decreases with the shell thickness due to the reduction of the local field magnitude  inside a thin metal shell.  We also found that the Landau damping rate exhibits quantum beats caused by the interference between  electron scattering paths from the nanoshell inner and outer metal-dielectric interfaces.


\acknowledgments
This work was supported in part by NSF Grant No. DMR-1610427 and No. HRD-1547754.

\appendix*
\section{}

Here we analyze function $G(\xi)$ in the interference term (\ref{Q12-final}). After shifting integration variables in Eq. (\ref{Q12-int}) as $\epsilon\rightarrow E_{F}+\epsilon-\omega_{L}/2$ and $\epsilon'\rightarrow E_{F}+\epsilon+\epsilon'$ and rescaling to $x=\epsilon/\omega_{L}$ and $s=l^{2}/(p_{F}R_{1})^{2}$, we arrive at (\ref{Q12-final}) with
%
\begin{align}
\label{G}
G(\xi)&=2\int_{-1/2}^{1/2}dx\int ds \sqrt{f(\xi,x,s)f(\xi,x,\kappa^{2} s)}
\nonumber\\ \times 
&\cos\left [\pi \omega_{L} \int_{-1/2}^{1/2}dx' \rho_{l}\bigl[E_{F}[1+\xi(x+x')]\bigr]\right],
\end{align}
where $f(\xi,x,s)=\sqrt{\left (1+\xi x -s\right )^{2}-\xi^{2}/4}$, and  the partial density of states is given by
\begin{align}
&\rho_{l}(\epsilon)=\frac{m}{\pi}\int_{R_{1}}^{R_{2}}\frac{dr}{p_{l}(\epsilon,r)}=\frac{R_{2}p_{l}(\epsilon,R_{2})-R_{1}p_{l}(\epsilon,R_{1}) }{2\pi\epsilon} 
\\
&= \dfrac{R_{2}}{\pi v_{F}} \left[ \dfrac{\sqrt{1+\xi (x+x^{\prime}) - \kappa^{2} s} - \kappa\sqrt{1+\xi (x+x^{\prime}) - s}}{ 1+\xi (x+x^{\prime})} \right].
\nonumber
\end{align}
For $\omega/E_{F}\ll 1$, the $x^{\prime}$-integrals are easily evaluated, yielding
\begin{align}
\label{G-close}
G(\xi)=2\int_{-1/2}^{1/2}dx & \int_{0}^{1+\xi x} ds \sqrt{f(\xi,x,s)f(\xi,x,\kappa^{2} s)}
\nonumber\\ \times
&\cos\left [w(\xi,x,s)\omega_{L}/\epsilon_{0}\right] ,
\end{align}
where $w(\xi,x,s)=\left(\sqrt{f(\xi,x,\kappa^{2}s)}-\kappa \sqrt{f(\xi,x,s)}\right)/(1+\xi x)$ with $f(\xi,x,s)\approx 1+\xi x -s$ (here $\epsilon_{0}= v_{F}/R_{2}$). Rescaling $s$ by $1+\xi x$, Eq.~(\ref{G-close}) factorizes as $G(\xi)=\int_{-1/2}^{1/2}dx \left (1+\xi x\right )^{2} S(\xi,x)$, where
\begin{align}
\label{G-simple}
S(\xi,x)=&2\int_{0}^{1}ds \sqrt{(1-s)(1-\kappa^{2} s)}
\nonumber\\ \times
&\cos\left [a(\xi,x)\left ( \sqrt{1-\kappa^{2} s} -\kappa\sqrt{1-s}\right )\right] ,
\end{align}
with shorthand notation $a(\xi,x)=(\omega_{L}/\epsilon_{0})/\sqrt{1+\xi x}$. With substitution $s=1-\frac{1-\kappa^2}{\kappa^2}\sinh^{2}\alpha$, $S$ is brought to the form
\begin{align}
\label{G-trig}
S(\xi,x)=\frac{4(1-\kappa^2)^2}{\kappa^3}&\int_{0}^{\alpha_{0}}d\alpha \left (\sinh\alpha\cosh\alpha\right )^{2} 
\nonumber\\ \times
&\cos\left [a(\xi,x)\sqrt{1-\kappa^{2}} e^{-\alpha}\right] ,
\end{align}
%
%
where $\sinh\alpha_{0}=\kappa/\sqrt{1-\kappa^{2}}$. For $a(\xi,x)\gg 1$, the integral is dominated by the upper limit, and for thin shells, $1-\kappa \ll 1$, corresponding to $\alpha_{0}>1$, can be evaluated as
\begin{equation}
\label{G-final}
S\approx -4 \frac{\sin(a\sqrt{1-\kappa^{2}} e^{-\alpha_{0}})}{a\sqrt{1-\kappa^{2}} e^{-\alpha_{0}}}
=  -4 \frac{\sin\left [a(1-\kappa)\right ]} {a(1-\kappa)}.
\end{equation}
With the above $S$ and after change of variable $t=\sqrt{1+\xi x}$, the expression for $G(\xi)$ takes the form
\begin{equation}
\label{G-factor}
G(\xi)=-\frac{8}{\xi D}\int_{t_{-}}^{t_{+}}dt t^{3}\sin (tD),
\end{equation}
where $t_{\pm}=\sqrt{1\pm\xi/2}$, and $D=(1-\kappa)\omega_{L}/\epsilon_{0}=\omega_{L}d/v_{F}$. Note that even though for $\xi\ll 1$ the integration interval is small, the integrand is still an oscillating function since $D\gg 1$ and so the product $D\xi$ can be arbitrary. In this case, a straightforward evaluation yields Eq. (\ref{G-anal}).


\end{document}